\documentclass[12pt]{JHEP3}

\usepackage{epsfig}
\epsfclipon
\usepackage{multicol}
\usepackage{epsfig,bm}
\usepackage{amssymb,amsmath}

\newcommand{\roughly}[1]{\mathrel{\raise.3ex\hbox{$#1$\kern-0.85em
\lower1ex\hbox{$\sim$}}}}

\newcommand{\lsim}{\roughly<}
\newcommand{\gsim}{\roughly>}
\newcommand{\sss}{\scriptscriptstyle}

\def\beq{\begin{equation}}
\def\eeq{\end{equation}}
\def\beqa{\begin{eqnarray}}
\def\eeqa{\end{eqnarray}}

\def\IK{\relax{\rm I\kern-.20em K}}
\def\IM{\relax{\rm I\kern-.15em M}}
\def\KKLMMT{{\IK L\IM T}}
\title{Warped Reheating in Brane-Antibrane Inflation}

\author{N.\ Barnaby,${}^1$ C.P.\ Burgess${}^{1,2,3}$ and J.M.\ Cline${}^1$\\
${}^1$ Physics Department, McGill University, 3600 University
Street\\
~~Montr{\'e}al, Qu{\'e}bec, Canada, H3A 2T8. \\
${}^2$ Department of Physics and Astronomy, McMaster University,\\
~~Hamilton, Ontario, Canada.\\
${}^3$ Perimeter Institute, Waterloo, Ontario, Canada. }

\abstract{We examine how reheating occurs after brane-antibrane
inflation in warped geometries, such as those which have recently
been considered for Type IIB string vacua. We adopt the standard
picture that the energy released by brane annihilation is dominantly
dumped into massive bulk (closed-string) modes which eventually
cascade down into massless particles, but argue that the this need
not mean that the result is mostly gravitons with negligible visible
radiation on the Standard Model brane.  We show that if the
inflationary throat is not too strongly warped, and if the string
coupling is sufficiently weak,  then a significant fraction of the
energy density from annihilation will be deposited on the Standard
Model brane, even if it is separated from the inflationary throat by
being in some more deeply warped throat. This is due to the 
exponential growth of the massive Kaluza-Klein wave functions toward
the infrared ends of the throats. We argue that the possibility of
this process removes a conceptual obstacle to the construction of
multi-throat models, wherein inflation occurs in a different throat
than the one in which the Standard Model brane resides. Such
multi-throat models are desirable because they can help to reconcile
the scale of inflation with the supersymmetry breaking scale on the
Standard Model brane, and because they can allow cosmic strings to be
sufficiently long-lived to be observable during the present epoch.}

\begin{document}

\makeatletter \@addtoreset{equation}{section} \makeatother
\renewcommand{\theequation}{\thesection.\arabic{equation}}

\setcounter{page}{1} \pagestyle{plain}
\renewcommand{\thefootnote}{\arabic{footnote}}
\setcounter{footnote}{0}


\section{Introduction}

There has been significant progress over the past years towards
the construction of {\it bona fide} string-theoretic models of
inflation. The main progress over early string-inspired
supergravity \cite{SugraInflation} and BPS-brane based
\cite{DvaliTye} models has come due to the recognition that
brane-antibrane \cite{BI1,BBbarInflation} and related
\cite{Angles,Others} systems can provide {\it calculable}
mechanisms for identifying potentially inflationary potentials.
Even better, they can suggest new observable signatures, such as
the natural generation of cosmic strings by the brane-antibrane
mechanism \cite{BI1,cosmicstrings}. The central problem to emerge
from these early studies was to understand how the many string
moduli get fixed, since such an understanding is a prerequisite
for a complete inflationary scenario.

Recent developments are based on current progress in
modulus stabilization within warped geometries with background
fluxes for Type IIB vacua \cite{GKP,Sethi,kklt}. Both
brane-antibrane inflation \cite{kklmmt} and modulus inflation
\cite{racetrackinflation} have been embedded into this context,
with an important role being played in each case by branes living
in strongly-warped `throat-like' regions within the extra
dimensions. These inflationary scenarios have generated
considerable activity \cite{OtherStringInflation,BIReviews}
because they open up the possibility of asking in a more focused
way how string theory might address the many issues which arise
when building inflationary models. For instance, one can more
fully compute the abundance and properties of any residual cosmic
strings which might survive into the present epoch
\cite{stringycosmicstrings}. Similarly, the possibility of having
quasi-realistic massless particle spectra in warped, fluxed Type
IIB vacua \cite{cmqu} opens up the possibility of locating where
the known elementary particles fit into the post-inflationary
world \cite{bcqs}, a prerequisite for any understanding of
reheating and the subsequent emergence of the Hot Big Bang.

Even at the present preliminary level of understanding, a
consistent phenomenological picture seems to require more
complicated models involving more than a single throat (in
addition to the orientifold images).\footnote{Two-throat models
are also considered for reasons different than those given here in
ref.~\cite{twothroat}.} This is mainly because for the
single-throat models the success of inflation and particle-physics
phenomenology place contradictory demands on the throat's warping.
They do so because the energy scale in the throat is typically
required to be of order $M_i \sim 10^{15}$ GeV to obtain
acceptably large temperature fluctuations in the CMB. But as was
found in ref.~\cite{bcqs}, this scale tends to give too large a
supersymmetry breaking scale for ordinary particles if the
Standard Model (SM) brane resides in the same throat. This problem
appears to be reasonably generic to the KKLT-type models discussed
to date, because these models tend to have supersymmetric anti-de
Sitter vacua until some sort of supersymmetry-breaking physics is
added to lift the vacuum energy to zero. The problem is that the
amount of supersymmetry-breaking required to zero the vacuum
energy also implies so large a gravitino mass that it threatens to
ruin the supersymmetric understanding of the low-energy
electroweak hierarchy problem.

No general no-go theorem exists, however, and there does appear to
be considerable room to try to address this issue through more
clever model-building. Ref.~\cite{RA} provides a first step in
this direction within the the framework of `racetrack' inflation
\cite{racetrackinflation}. Another possibility is a picture having
two (or more) throats, with inflation arising because of
brane-antibrane motion in one throat but with the Standard Model
situated in the other (more about this proposal below). By
separating the scales associated with the SM and inflationary
branes in this way, it may be possible to reconcile the
inflationary and supersymmetry-breaking scales with one another.

Besides possibly helping to resolve this problem of scales,
multi-throat models could also help ensure that string defects
formed at the end of inflation in the inflationary throat have a
chance of surviving into the present epoch and giving rise to new
observable effects \cite{stringycosmicstrings}. They are able to
do so because if the Standard Model were on a brane within the
same throat as the inflationary branes, these defects typically
break up and disappear by intersecting with the SM brane.

At first sight, however, any multi-throat scenario seems likely to
immediately founder on the rock of reheating.\footnote{See
ref.~\cite{braneheat} for a discussion of issues concerning
brane-related reheating within other contexts.} Given the absence
of direct couplings between the SM and inflationary branes, and
the energy barrier produced by the warping of the bulk separating
the two throats, one might expect the likely endpoint of
brane-antibrane annihilation to be dump energy only into
closed-string, bulk modes, such as gravitons, rather than visible
degrees of freedom on our brane. In such a universe the energy
which drove inflation could be converted almost entirely into
gravitons, leaving our observable universe out in the cold.

It is the purpose of the present work to argue that this picture
is too pessimistic, because strongly-warped geometries provide a
generic mechanism for channelling the post-inflationary energy
into massless modes localized on the throat having the strongest
warping. They can do so because the massive bulk Kaluza-Klein (KK)
modes produced by brane-antibrane annihilation prefer to decay
into massless particles which are localized on branes within
strongly-warped throats rather than to decay to massless bulk
modes. As such, they open a window for obtaining acceptable
reheating from brane-antibrane inflation, even if the inflationary
and SM branes are well separated on different throats within the
extra dimensions.

The remainder of the paper is organized as follows. In \S2 we
introduce a simple generalization of the Randall-Sundrum (RS)
model \cite{RS} containing two AdS$_5$ throats with different warp
factors, as a tractable model for the \KKLMMT\ inflationary
scenario \cite{kklmmt} with two throats.  Here we recall the form
of the KK graviton wave functions in the extra dimension. This is
followed in \S3 by an account of how the tachyonic fluid
describing the unstable brane-antibrane decays into excited
closed-string states, which quickly decay into KK gravitons.
\S4 Discusses the tunneling of the KK modes
through the energy barrier which exists between the
two throats because of the warped geometry. 
 \S5
Gives an estimate of the reheating temperature on the SM brane
which results from the preferential decay of the KK gravitons into
SM particles. Our conclusions are given in \S6.

\section{Tale of Two Throats}

We wish to describe reheating in a situation where brane-antibrane
inflation occurs within an inflationary throat having an energy
scale of $M_i$, due to the warp factor $a_i = M_i/M_p$, where
$M_p$ is the 4D Planck mass. This throat is assumed to be
separated from other, more strongly warped, throats by a weakly
warped Giddings-Kachru-Polchinski (GKP) manifold \cite{GKP} whose
volume is only moderately larger than the string scale, so $M_s
\lsim M_p$. In the simplest situation there are only two throats
(plus their orientifold images), with the non-inflationary
(Standard Model) throat having warp factor $a_{sm} \ll a_i$.

There are two natural choices for the SM warp factor, depending on
whether or not the SM brane strongly breaks 4D supersymmetry. For
instance, if the SM resides on an anti-D3 brane then supersymmetry
is badly broken and the SM warp factor must describe the
electroweak hierarchy {\it \`a la} Randall and Sundrum \cite{RS}, with
$a_{sm} \sim M_W/M_p \sim 10^{-16}$. Alternatively, if the SM
resides on a D3 or D7 brane which preserves the bulk's $N=1$
supersymmetry in 4D, then SUSY breaking on the SM brane is
naturally suppressed by powers of $1/M_p$ because it is only
mediated by virtual effects involving other SUSY-breaking anti-D3
branes. In this case the electroweak hierarchy might instead be
described by an intermediate-scale scenario \cite{IntScale}, where
$a_{sm} \sim M_{int}/M_p \sim (M_W/M_p)^{1/2} \sim 10^{-8}$.

A potential problem arises with the low-energy field theory
approximation if $a_{sm} < a_i^2$, because in this case the string
scale in the SM throat, $M_{sm} \sim a_{sm} M_p$, is smaller than
the inflationary Hubble scale $H_i \sim M_i^2/M_p \sim a_i^2 M_p$
\cite{Rob}. In this case string physics is expected to become
important in the SM throat, and stringy corrections may
change the low-energy description. The intermediate
scale, where $a_{sm} \gsim a_i^2$, 
 is more attractive from this point of view, since
for it the field-theory approximation may be justified.

To proceed we use the fact that within the GKP compactification the
geometry within the throat is well approximated by
\beq \label{RS}
    ds^2 = a^2(y)(dt^2 - dx^2) - dy^2 - y^2 \, d\Omega^2_5 \,,
\eeq
where $y$ represents the proper distance along the throat, $a(y) =
e^{-k|y|}$ is the throat's warp factor and $d\Omega^2_5$ is the
metric on the base space of the corresponding conifold singularity
of the underlying Calabi-Yau space \cite{throatmetric}. Of most
interest is the 5D metric built from the observable 4 dimensions
and $y$, which is well approximated by the metric of 5-dimensional
anti-de Sitter space.

A simple model of the two-throat situation then consists of
placing inflationary brane-antibranes in a throat at $y=-y_i$ and
putting the Standard Model brane at $y=+y_{sm}$, as is illustrated
in Fig.~\ref{fig1}. Our analysis of this geometry follows the
spirit of ref.~\cite{dkkls}. Since most of the interest is in the
throats, we simplify the description of the intervening bulk
geometry by replacing them with a Planck brane at $y=0$, with the
resulting discontinuity in the derivative of $a(y)$ chosen to
reproduce the smoother (but otherwise similar) change due to the
weakly-warped bulk. This approximation is illustrated in
Fig.~\ref{fig2}, with the smooth dashed curve representing the
warp factor in the real bulk geometry and the solid spiked curve
representing the result using an intervening Planck brane instead.

\DOUBLEFIGURE[ht]{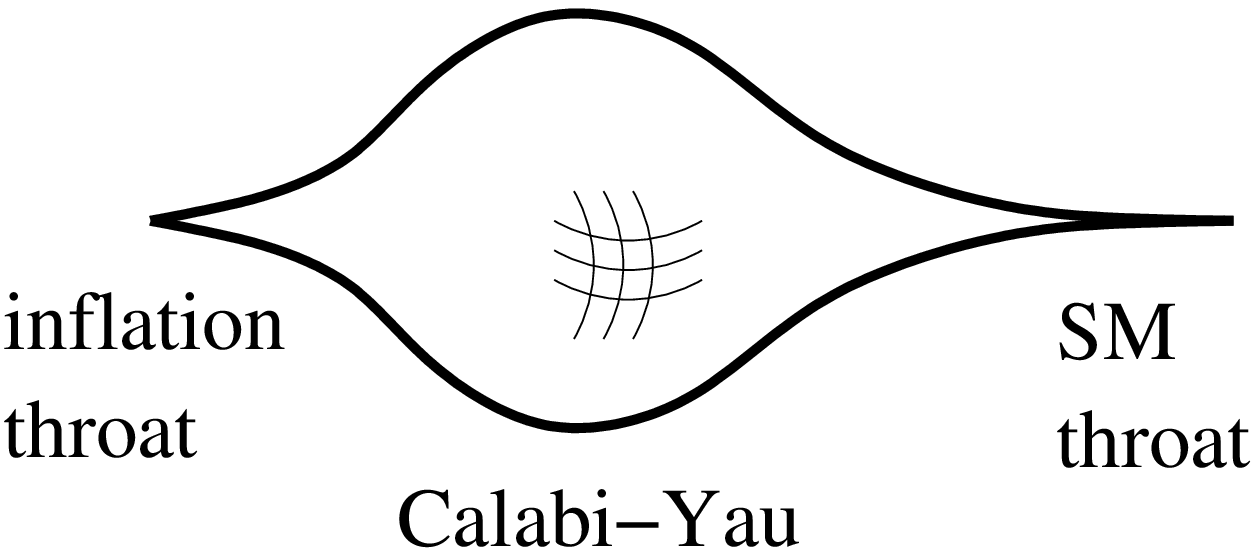, width=1.2\hsize} {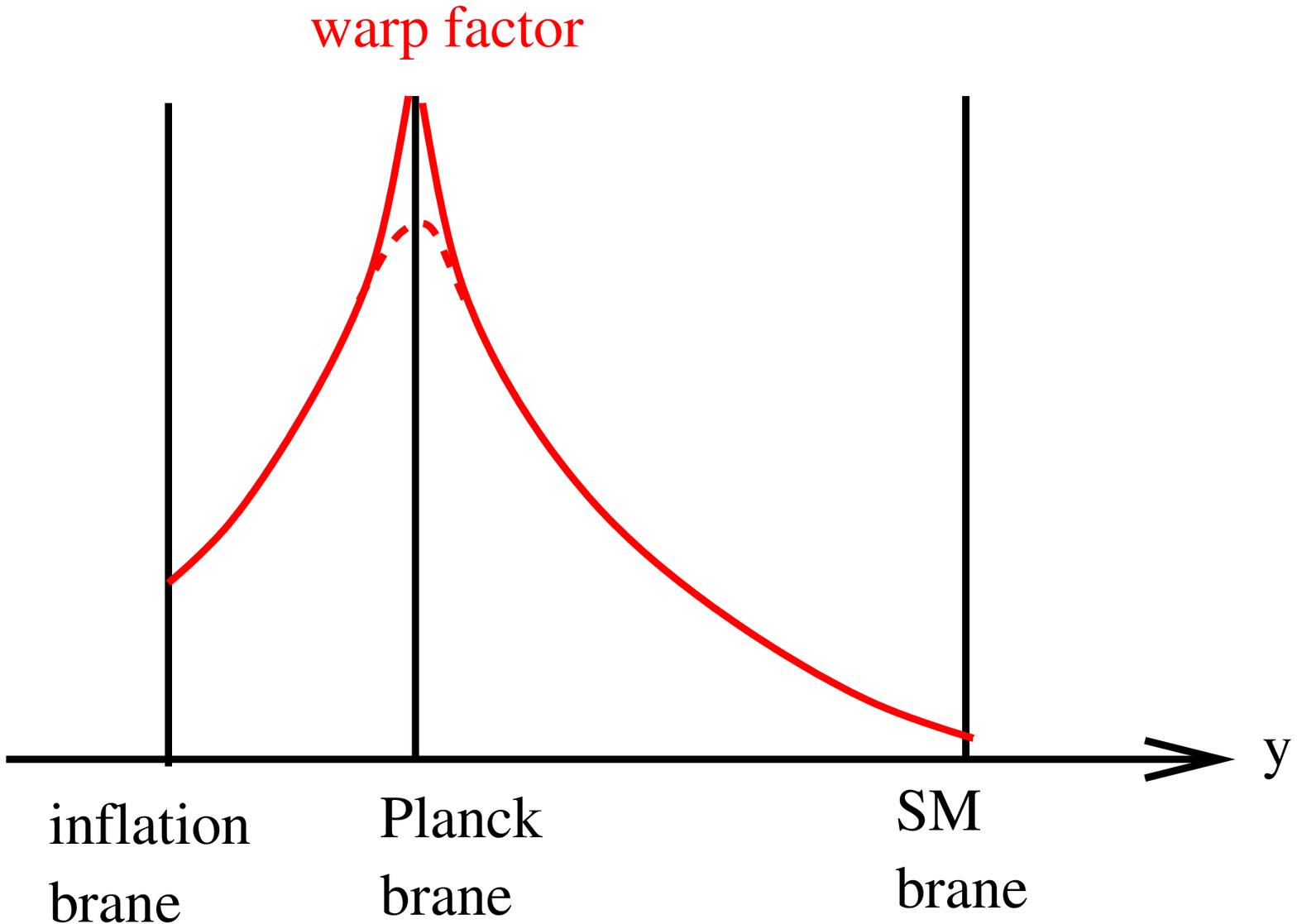,
width=1.2  \hsize}{A Type IIB vacuum with a mildly warped
inflationary throat and a strongly warped Standard Model throat.
This diagram suppresses any image throats arising due to any
orientifolds which appear in the compactification.\label{fig1}}
{The warp factor as function of a bulk radial coordinate in a
simplified model of two asymmetric throats. As shown in the
figure, the part of the internal space outside of the throats can
be regarded as a regularization of a `Planck' brane of a
Randall-Sundrum geometry.\label{fig2}}

Of particular interest in what follows are the massive
Kaluza-Klein modes in the bulk, since these are arguably the most
abundantly-produced modes after brane-antibrane annihilation. For
instance, focussing on the 5 dimensions which resemble AdS space
in the throat, a representative set of metric fluctuations can be
parameterized as $h(x,y)$ in the line element,
\beq \label{RSh}
    ds^2 = a^2(y)(dt^2 - dx^2 + h_{\mu\nu} dx^\mu dx^\nu) - dy^2
    \,.
\eeq
In the static AdS background, the KK modes have spatial
wavefunctions of the form
\beq
    h(x,y) = \sum_n\phi_n(y) \; e^{ip\cdot x}
\eeq
with $p\cdot x = - E_n t + {\bf p} \cdot {\bf x}$, and $\phi_n(y)$
satisfying the equation of motion
\beq
    -{d\over dy}\left(e^{-4k|y|} \,{d\phi_n\over dy}\right)
    = m^2_n e^{-2k|y|}\phi_n \,.
\eeq
Here $m_n^2 = p \cdot p$ is the mode's 4D mass as viewed by
brane-bound observers.

Exact solutions for $\phi_n(y)$ are possible in the Planck-brane
approximation \cite{RS,dkkls,PlanckBraneModes}, and are linear
combinations of Bessel functions times an exponential
\beq \label{modefunctions}
    \phi_n(y) = N_n \, e^{2k|y|}
    \left[ J_2\left({m_n\over k}e^{k|y|}\right)
    + {b_n} \,
    Y_2\left({m_n\over k}e^{k|y|}\right) \right]
 \,
\eeq
where, for low lying KK modes ($m_n \ll k$) one has
\beq
    b_n \cong \frac{\pi m_n^2}{4 k^2}
\eeq
while for heavy KK modes ($m_n / k \cong 1$) one has
\beq
    b_n \cong -0.47 + 1.04 \left( \frac{m_n}{k} \right).
\eeq
$N_n$ is determined by the orthonormality condition, which ensures
that the kinetic terms of the KK modes are independent of
$a_{sm}$:
\beq
    \int_{-y_i}^{y_{sm}} dy\, e^{-2k|y|} \phi_n \phi_m =
    \delta_{nm}\,.
\eeq
These wavefunctions are graphed schematically in Fig.~\ref{fig3}.

For strongly-warped throats it is the exponential dependence which
is most important for the KK modes. Because of the exponential
arguments of the Bessel functions in eq.~(\ref{modefunctions}),
the presence of the Bessel functions modifies the large-$y$
behaviour slightly. Due to the asymptotic forms $J_2(z) \propto
z^{-1/2}$ for large $|z|$, and similarly for $Y_2(z)$, we
see that $\phi_n(y) \sim e^{3k|y|/2}$ for $m_n e^{k|y|} \gg k$. It
is only this behaviour which we follow from here on. Taking the
most warping to occur in the SM throat we find that $m_n$ is
approximately quantized in units of $M_{sm} \equiv a_{sm} M_p$,
which is either of order $M_W \sim 10^3$ GeV or $M_{int} \sim
10^{10}$ GeV depending on whether or not supersymmetry breaks on
the SM brane. Keeping only the exponentials we find that
orthonormality requires $N_n^{-2} \sim \int^{y_{sm}} dy \;
e^{-2ky} \, \left( e^{3k |y|/2} \right)^2 \sim (k \,
a_{sm})^{-1}$, and so
\beq
\label{lownwf}
    \phi_n(y) \sim  ({a_{sm}\, k})^{1/2}\, e^{3k|y|/2} \,,
\eeq
showing that these modes are strongly peaked deep within the
throat. This is intuitively easy to understand, since being
localized near the most highly-warped region allows them to
minimize their energy most effectively.

Thus, even the most energetic KK modes still have exponentially
larger wave functions on the TeV brane, with the more energetic
modes reaching the asymptotic region for smaller $y$. This is
illustrated in Figure \ref{fig5}, which shows $\ln|\phi_n(y)|$
versus $y$ in the representative case of a throat having warp
factor $a = e^{-10}$, for a series of KK states with masses going
as high as $M_p$ ($n=20,000$). As the figure shows, the wave
functions grow exponentially toward the TeV brane, with the onset
of the asymptotic exponential form setting in earlier for larger
mode number.\footnote{For the lowest-lying modes having the
smallest nonzero masses it can happen that the asymptotic form of
the Bessel functions is not yet reached even when $y = y_{sm}$, in
which case the exponential peaking is slightly stronger than
discussed above.} This behaviour is central to the estimates which
follow.

Among the KK modes it is the zero modes which are the exceptional
case because their wavefunction is constant, $\phi_0 \sim
\sqrt{k}$, and so they are not exponentially peaked inside the
throat. It is the strong exponential peaking of the lightest
massive KK modes relative to the massless modes which is central
to the reheating arguments which follow.

\DOUBLEFIGURE[ht]{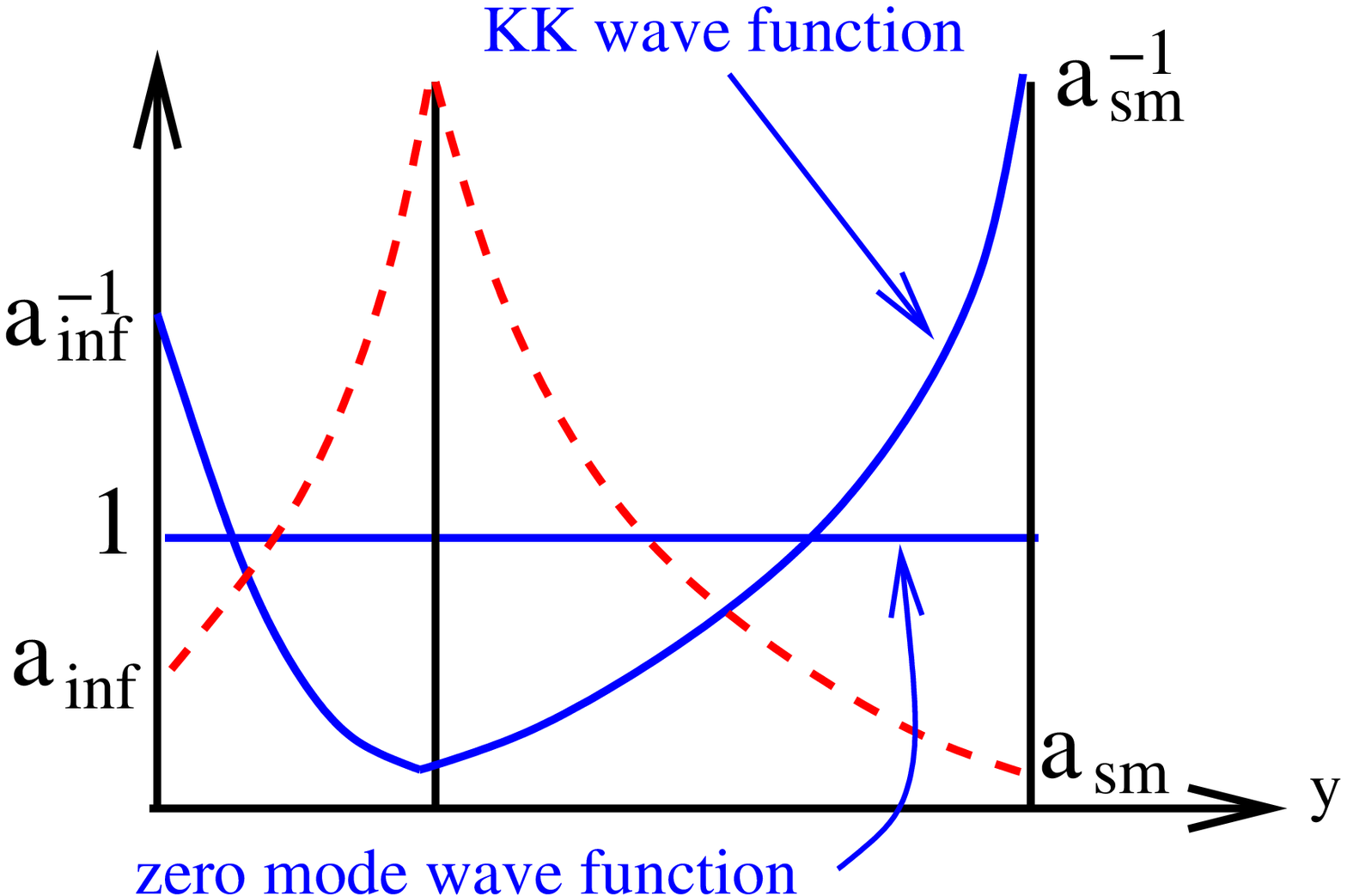, width=6cm}{wf3.eps,
width=6cm}{Wave functions of KK gravitons on the internal
space.\label{fig3}}{Unnormalized wave functions for highly excited
KK gravitons with KK numbers $n=1$, 100, 1000 and
20,000.\label{fig5}}

\EPSFIGURE[ht]{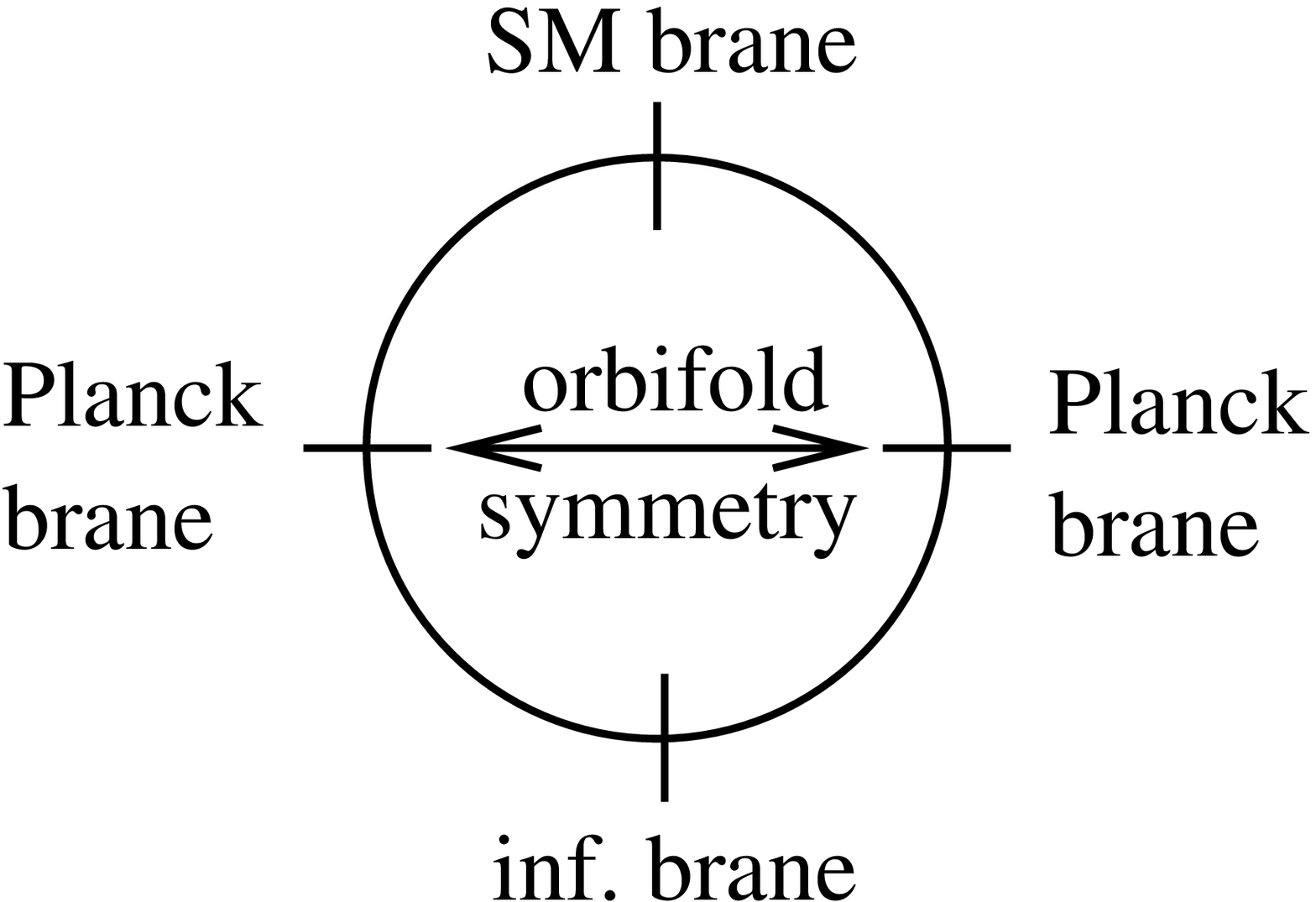, width=6cm}{How to place two throats
on an $S_1$ with $Z_2$ orbifold symmetry.\label{fig4}}

In our simplified model, the presence of two throats is not much
more complicated than the original RS model.  Mathematically it is
the same, except that RS identified the two sides
$y\leftrightarrow -y$ through orbifolding.  Instead we interpret
them as two separate throats with different depths defined by the
brane locations $-y_i$ and $+y_{sm}$.  One can imagine doubling
this entire system on $S_1$ and orbifolding as shown in
Fig.~\ref{fig4}, to define the boundary conditions on the metric
and its perturbations at the infrared branes. In this figure, the
orbifold identification acts horizontally so that the inflation
and SM branes are distinct fixed points.

\section{Brane-Antibrane Annihilation}

In brane-antibrane inflation the energy released during reheating
is provided by the tensions of the annihilating branes. Although
this annihilation process is not yet completely understood,
present understanding indicates that the energy released passes
through an intermediate stage involving very highly-excited string
states, before generically being transferred into massless
closed-string modes. The time frame for this process is expected
to be the local string scale.

For instance many of the features of brane-antibrane annihilation
are believed to be captured by the dynamics of the open-string
tachyon which emerges for small separations for those strings that
stretch between the annihilating branes.\footnote{See, however,
ref.~\cite{BraneRad} for a discussion of an alternative mechanism
for which the relevant highly-excited strings are open strings,
but for which the annihilation energy nonetheless eventually ends
up in massless closed string modes.} In flat space and at zero
string coupling ($g_s=0$), the annihilation instability has been
argued to be described by the following tachyon Lagrangian
\cite{Sen}
\beq \label{Sen}
    {\cal L}_T = -2\tau_0\, e^{-|T|^2/l_s^2} \sqrt{1-|\partial_\mu T|^2}
\eeq
where $T$ is the complex tachyon field, $\tau_0$ is the tension of
either of the branes, and $l_s$ is the string length scale. During
inflation, when the brane and anti-brane are well separated, $\dot
T = 0$ and the pressure of the system $p_i$ is simply the negative
of the tension of the two branes, $p_i=-\rho_i$, while afterward
$\dot T\to 1$ and $p_i\to 0$. In this description the pressureless
tachyonic fluid would dominate the energy density of the universe
and lead to no reheating whatsoever.

However, for nonvanishing $g_s$, the time evolution of the tachyon
fluid instead very quickly generates highly excited closed-string
states \cite{BraneDecay,Sen-review}. For D$p$ systems with $p>2$
the rate of closed string production in this process is formally
finite, whereas it diverges for $p\le 2$ (and so passes beyond the
domain of validity of the calculation). This divergence is
interpreted to mean that for branes with $p\le 2$ all of the
energy liberated from the initial brane tensions goes very
efficiently into closed string modes. For spatially homogeneous
branes with $p>2$ the conversion is less efficient and so can be
dominated by other, faster processes. In particular, it is
believed that these higher-dimensional branes will decay more
efficiently inhomogeneously, since they can then take advantage of
the more efficient channels which are available to the
lower-dimensional branes. For example a D3 brane could be regarded
as a collection of densely packed but smeared-out D0 branes, each
of which decays very efficiently into closed strings. Since the
decay time is of order the local string scale, $l_s = 1/M_s$, the
causally-connected regions in this kind of decay are only of order
$l_s$ in size, and so have a total energy of order the brane
tension times the string volume, $\tau_0 l_s^3 \sim M_s/g_s$.

These flat-space calculations also provide the distribution of
closed-string states as a function of their energy. The energy
density deposited by annihilating D3-branes into any given string
level is of order $M_s^4$, and so due to the exponentially large
density of excited string states the total energy density produced
is dominated by the most highly-excited states into which decays
are possible. Since the available energy density goes like $1/g_s$
the typical closed-string state produced turns out to have a mass
of order $M_s/g_s$, corresponding to string mode numbers of order
$N \sim 1/g_s$. On the other hand, the momentum transverse to the
decaying branes for these states turns out to be relatively small,
$p_{\sss T} \sim M_s/\sqrt{g_s}$ \cite{BraneDecay,Sen-review}, and
so the most abundantly produced closed-string states are
nonrelativistic.

How do these flat-space conclusions generalize to the warped Type
IIB geometries which arise in string inflationary models? If the
annihilating 3-branes are localized in the inflationary throat,
then the tension of the annihilating branes is of order $\tau_0
\sim (a_i M_s)^4/e^{\phi_i}$, where $\phi_i$ denotes the value
taken by the dilaton field at the throat's tip. The highly-excited
closed-string states that are produced in this way live in the
bulk, with the energy density produced being dominated by those
whose masses are of order $a_i M_s/e^{\phi_i}$. Once produced,
these closed-string bulk modes decay down to lower energies and,
as might be expected from phase space arguments, most of them
typically drop down to massless string states very quickly. An
important exception to this would arise for those states carrying
the most angular momentum at any given string mass level, since
these must cascade more slowly down to lower energies in order to
lose their angular momentum \cite{HiJCascade}. However these seem
unlikely to be produced in appreciable numbers by brane-antibrane
annihilation.

We are led in this way to expect that the annihilation energy is
distributed relatively quickly amongst massless string states, or
equivalently to KK modes of the higher-dimensional supergravity
which describes these states. Although the initial massive string
modes would be nonrelativistic, with $M \sim a_i M_s/e^{\phi_i}$
and $p_{\sss T} \sim a_i M_s/e^{\phi_i/2}$, the same need not be
true for the secondary string states produced by their decay,
whose masses are now of order the KK mass scales. Consequently
these states may be expected not to remain localized in the
inflationary throat, and so if the extra dimensions are not too
large compared with the string scale these modes would have time
to move to the vicinity of the SM throat before decaying further.
Once there, they would be free to fall into the potential wells
formed by the throats as their energy is lost by subsequent decays
into lower-energy levels.

This physical picture is supported by the exponential peaking of
the KK-mode wave-functions in the most deeply-warped throats. In
order to estimate the efficiency with which energy can be
transferred amongst KK modes, we can use the approximate behavior
of the wave functions given in the previous section to keep track
of powers of the throat's warp factor, $a_{sm}$. For instance,
consider the trilinear vertex among 3 KK states having mode
numbers $n_1$, $n_2$ and $n_3$ which is obtained by dimensionally
reducing the higher-dimensional Einstein-Hilbert action,
$\sqrt{g}R$. Keeping in mind that $\sqrt{g} g^{\mu\nu} \propto
a^2$ and that $\psi_n \propto a_{sm}^{1/2}/a^{3/2}$ for the
nonzero modes (eq.\ (\ref{lownwf})), we find that the trilinear
vertex involving $0 \le r \le 3$ massive KK modes (and $3-r$
massless KK modes) has the following representative estimate
\beqa
    {\cal L}_{\rm int} &\sim& \int^{y_{sm}}_{-y_i} dy \, \sqrt{g}
    \, g^{\mu\nu} g^{\alpha\beta} g^{\kappa\sigma} g^{\rho\delta}
    h_{\alpha\kappa} \partial_\mu h_{\sigma\rho} \partial_\nu
    h_{\beta\delta} \nonumber \\
    &\sim& \int^{y_{sm}}_{-y_i} dy\, e^{-2k|y|}\, \eta^{\mu\nu}
    \psi_{n_1}(x,y)\,
    \partial_\mu \psi_{n_2}(x,y)\, \partial_\nu
    \psi_{n_3}(x,y) \nonumber\\
    &\sim&   \psi_{n_1}(x)\, \psi_{n_2}(x)\, \psi_{n_3}(x)\,
     p_2\!\cdot\!p_3\; a_{sm}^{r/2} \,
     \int^{y_{sm}}_{-y_i} dy\, e^{-2k|y|}
     (e^{3k|y|/2})^r \nonumber\\
    &\sim& \psi_{n_1}(x)\,
    \psi_{n_2}(x)\,\psi_{n_3}(x)\, \left(
    \frac{p_2\!\cdot\!p_3}{k}
    \right) \; a_{sm}^{\eta} \,
\eeqa
where
\beq
    \eta = 2-r \qquad \hbox{if} \quad r \ge 2, \qquad \hbox{and}
    \qquad \eta = \frac{r}{2} \qquad \hbox{if} \quad r = 0,1  \,.
\eeq
Notice for this estimate that since derivatives in the
compactified directions are proportional to $g^{mn}$ rather than
$g^{\mu\nu}$, they suffer from additional suppression by powers of
$a = e^{-ky}$ within the throat. Here $m$, $n$ label the internal
directions perpendicular to the large $3+1$-dimensional Minkowski
space.

Thus a trilinear interaction amongst generic KK modes ($r=3$),
even those with very large $n$, is proportional to $1/(a_{sm} k)
\sim 1/M_{sm}$, and so is only suppressed by inverse powers of the
low scale. Similarly, $r=2$ processes involving two massive KK
modes $B$ and $B'$, and one massless bulk mode $ZM$ --- such as
the reaction $B \to B' + ZM$ --- are $\propto 1/k \sim 1/M_p$ and
so have the strength of 4D gravity inasmuch as they are Planck
suppressed. The same is also true of the $r=0$ couplings which
purely couple the zero modes amongst themselves.\footnote{The
appendix shows that this agrees with the size of the couplings
found in the effective 4D supergravity lagrangian which describes
the zero-mode and brane couplings.} Finally, those couplings
involving only a single low-lying massive mode and two zero modes
($r = 1$) --- such as for $B \to ZM + ZM'$
--- are proportional to $a_{sm}^{1/2}/k \sim (M_{sm}/M_p^3)^{1/2}$
and so are even weaker than Planck-suppressed.

Similar estimates may also be made for the couplings of the
generic and the massless KK modes to degrees of freedom on a brane
sitting deep within the most strongly-warped throat. Using the
expressions $\phi_0(y_{sm}) \sim 1$ and $\phi_n(y_{sm}) \sim
1/a_{sm}$ for massless and massive KK modes respectively, this
gives:
\beqa
    {\cal L}_i &=& M_p^{-1}\left({h_{\mu\nu}^{(0)}\phi_0(y_{sm})} +
    \sum_n{h_{\mu\nu}^{(n)}\phi_n(y_{sm})}\right)
    T^{\mu\nu}_{sm} \nonumber\\
     &\sim& \left({h_{\mu\nu}^{(0)}\over M_p} +
    \sum_n{h_{\mu\nu}^{(n)}\over M_{sm}}\right) T^{\mu\nu}_{sm}
    \,.
\eeqa
We see here the standard Planck-suppressed couplings of the
massless modes (such as the graviton) as compared with the
$O(1/M_{sm})$ couplings of the massive KK modes.

The picture which emerges is one for which the energy released by
brane-antibrane annihilation ends up distributed among the massive
KK modes of the massless string states. Because the wavefunctions
of these modes tend to pile up at the tip of the most warped (SM)
throat, their couplings amongst themselves --- and their couplings
with states localized on branes in this throat --- are set by the
low scale $M_{sm}$ rather than by $M_p$. Furthermore, because the
${\cal O}(1/M_{sm})$ couplings to the massless modes on the SM
branes are much stronger than the Planck-suppressed couplings to
the massless bulk modes, we see that the ultimate decay of these
massive KK modes is likely to be into brane states.   If it were not
for the issue of tunneling, which we consider below, the final
production of {\it massless} KK zero modes would be highly suppressed.
Although we make the argument here for gravitons, the
same warp-counting applies equally well to the other fields
describing the massless closed-string sector.

In summary, we see that strong warping can provide a mechanism for
dumping much of the energy released by the decay of the unstable
brane-antibrane system into massless modes localized on branes
localized at the most strongly-warped throat, regardless of
whether the initial brane-antibrane annihilation is located in
this throat. It does so because the primary daughter states
produced by the decaying brane-antibrane system are expected to be
very energetic closed strings, which in turn rapidly decay into
massive KK modes of the massless string levels. The strong warping
then generically channels the decay energy into massless modes
which are localized within the most strongly-warped throats,
rather than into massless bulk modes.

\section{Tunneling}
However the above arguments are too naive, since they ignore the fact that
there is an energy barrier which the initial KK gravitons must tunnel
through in order to reach the Standard Model throat.  The efficiency
of reheating on the SM brane will be suppressed by the tunneling
probability.

The tunneling amplitude for a KK mode with energy $E_n$, in a
Randall-Sundrum-like two-throat model just like ours,
has been computed exactly in \cite{dkkls}:
\beq	
\label{tunnel}
{\cal A} 
\sim a_{\rm inf}^2 \left({E_n\over M_i}\right)^2 
\eeq
where $M_i$ is the characteristic energy scale at the bottom of the
inflation throat, out of which the particle is tunneling.
Intuitively, this can be understood in the following way.
For a mode with minimum (but nonzero) energy,  
the tunneling amplitude is given by
the ratio of its wave function at the bottom of the throat to that at
the top:
\beq	{\cal A} \sim {\phi_n({y_{\rm inf}})\over \phi_n(0)}
\sim a_{\rm inf}^2 \eeq
Since energies in the throat scale linearly with the warp factor,
a high-energy mode, with energy $E_n$, should have the larger tunneling
amplitude given by (\ref{tunnel}).
In the present case, the highest KK modes have
energies determined by the tension of a D0-brane (as argued above);
but we must remember that it is the warped tension which counts, so
the maximum energy scale is given by
$E_n \sim a_{\rm inf}\, {M_s/g_s}$
whereas the characteristic energy scale in the throat is $M_i \sim
a_{\rm inf}\,M_s$.  The tunneling probability is therefore
\beq
\label{prob}
	P = {\cal A}^2 = \left({a_{\rm inf}\over g_s}\right)^4	\eeq
To maximize this, we need a high scale of inflation (so that the 
inflationary warp factor is not too small) and a small string
coupling.  Optimistically, we could imagine that inflation is taking
place near the GUT scale, $10^{16}$ GeV, which saturates the bound
on the inflation scale coming from gravitational waves contributing to
the CMB anisotropy, and $g_s = 0.01$.  Then $a_{\rm inf} = 10^{-3}$
and the tunneling probability is $P=10^{-4}$.

With a small tunneling probability $P$, the universe immediately
after reheating would be dominated by massless gravitons, the final
decay product of KK gravitons confined to the inflation throat. Only
the small fraction $P$ of the original false vacuum energy density 
which tunneled into the SM throat would efficiently decay into
ultimately visible matter on the SM brane.  Such a distribution of
energy density would be strongly ruled out by big bang
nucleosynthesis were it to persist down to low temperatures.  There
are several natural ways in which this outcome can be avoided
however.  Since reheating occurs at a high scale (given that $P$ is
not {\it too} small, as we shall quantify in the next section),  the
number of effectively massless degrees of freedom $N(T_{\rm rh})$ could be
quite large at the temperature of reheating.  As the heavier of these
species go out of equilibrium, they transfer their entropy into the
lighter visible sector particles, resulting in a relative enhancement
factor $N(T_{\rm rh})/N(T_{\rm nuc})$ of the entropy in visible radiation at
the nucleosynthesis temperature $T_{\rm nuc}$.  On the other hand
the entropy density in gravitons remains fixed because they
were already thermally decoupled from the moment they were produced.
If this is the only mechanism for diluting gravitons, we would require
$P\,N(T_{\rm rh})/N(T_{\rm nuc}) \gsim 10$, so that gravitons make up no more 
than 10\% of the total energy density at BBN.

Additionally, gravitons can be efficiently diluted if any heavy
particles decay out of equilibrium at a temperature $T_{\rm dec}$
before BBN, so that they come to
dominate the energy density during a significant interval.\footnote{We
thank Andrei Linde for pointing out this possibility.}\  In this case
the gravitons are diluted by an additional factor of $T_{\rm dec}/M$ by
decaying particles of mass $M$.  Similarly, a period of domination by
coherent oscillations of a scalar field (for example a flat direction
with a large initial VEV, that gets lifted during a phase transition)
will behave as though matter-dominated, and give the same kind of
dilution.

There is one more criterion which must be satisfied in order for
tunneling to be significant: the lifetime of the heaviest KK states
should be of the same order as or longer than the typical tunneling
time. The typical momentum of the KK modes is of order $M_i/\sqrt{g_s}$,
hence the velocity transverse to the decaying brane is of order
$\sqrt{g_s}$
\cite{BraneDecay}, and the length of the throat is of order
$R=cM_s^{-1}$ where $c\gsim 10$ in order to have a reliable
low-energy description of the inflationary dynamics.  Thus the
tunneling time is
\beq	\tau_{t} \gsim 10{ M_s^{-1}\over \sqrt{g_s}} P^{-1} \eeq
while the lifetime of a KK mode is estimated to be
\beq
	\tau_{l} \sim \left(g_s^2\,{M_i\over g_s}\right)^{-1} \eeq
where the factor of $g_s^2$ comes from the squared amplitude for the two-body
decay and $M_i/g_s$ is the phase space. 
This translates into the  requirement $10\,g_s^{9/2} <
a_{\rm inf}^3$.  To satisfy this, we need to take a string coupling
which is somewhat smaller than $0.01$, say $g_s = 0.006$.  The
bound is then saturated for an inflationary warp factor of $a_{\rm
inf} = 10^{-3}$.  In this case the tunneling probability is
$8\times 10^{-4}$.

It would be interesting if there exist warped compactifications
in which the background dilaton field is varying between the two
throats.  In this case it may be possible to have a smaller string
coupling in the inflation throat, as would be desirable for the
tunneling problem, while keeping the string coupling in the SM throat
at a phenomenologically preferred value.

It is worth emphasizing that even when the tunneling probability
for energetic KK modes is large enough for reheating, the lifetime
of cosmic strings in the inflation throat is still cosmologically
large.  Copeland {\it et al.} \cite{stringycosmicstrings} 
estimate the barrier penetration amplitude for
a string to be
\beq	e^{-1/a_{\rm inf}^2} \sim e^{-10^{6}} \eeq
which makes the strings stable on cosmological time scales.

\section{Warped Reheating}

{}From the previous sections we see that the endpoint of
brane-antibrane inflation can be considered as a gas of 
nonrelativistic closed strings with mass $M_i/g_s$, density $M_i^3$
and decay rate $\Gamma \sim g_s\,M_i$, localized in the inflationary
throat.  These heavy states cascade
down to massless gravitons through a sequence of KK gravitons,
a fraction $P$ (eq.\ \ref{prob}) of which tunnel to the SM brane
and decay into visible sector particle.

Initially there will be two relevant reheat temperatures: one 
for the massless gravitons, $T_{\rm grav}$ and one for the 
visible sector, $T_{\rm vis}$.
By the standard reheating estimate \cite{Reheat} we see that 

\beq
\label{reheat-grav}
    T_{\rm grav} \sim 0.1 \, (\Gamma M_p)^{1/2} \sim 0.1 \,
    a_i^{1/2} \, \sqrt{M_s\, M_p} \sim 10^{-3} \, M_p \,,
\eeq
where the last estimate uses $a_i \sim 10^{-3}$. and $M_s\sim M_p/10$.
On the other hand, since a fraction $P$ of the false vacuum energy 
was converted to visible sector particles, we deduce that

\beq
\label{reheat-vis}
    T_{\rm vis} = P^{1/4}\, T_{\rm grav} \sim T_{\rm grav}/10
\eeq
using the optimistic estimate of the previous section for $P$. This
estimate is high enough to avoid potential problems to which a low
reheat temperature can give rise.  One should take this result with a
grain of salt since it   is marginally larger than both $M_i$ and
$M_{sm}$, and because it is larger than the string scale in the
throats it invalidates the 4D field-theoretic calculation on which it
is based. A more careful calculation must instead be based on a
higher-dimensional, string-theoretic estimate of the energy loss,
which goes beyond the scope of this article.

In conventional inflation models, such a high reheating temperature would
be in conflict with the gravitino bound (overproduction of gravitinos,
whose late decays disrupt big bang nucleosythesis).  It is interesting in
this regard that the KKLT scenario gives a very large gravitino mass,
around $m_{3/2} =  6\times 10^{10}$ GeV \cite{RA}, which is so large that
there is effectively no upper limit on the reheat temperature (see for
instance ref.\ \cite{KKM}).  The disadvantage of such a large gravitino
mass is that supersymmetry is broken at too high a scale to explain the
weak scale of the SM. If SUSY is this badly broken, one possibility for
explaining the weak hierarchy is that the large landscape of string vacua
provides a finely-tuned Higgs mass, as well as cosmological constant, as
has been suggested in ref.\ \cite{splitsusy}.  If this is the case, then
the degree of warping in the SM model brane would not be crucial for
determining the TeV scale, and the existence of an extra throat to
contain the SM model brane would be unnecessary.  However, given the
large number of 3-cycles in a typical Calabi-Yau manifold,  each of which
can carry nontrivial fluxes, the existence of many throats should be
quite generic, and it would not be surprising to find the SM brane in a
different throat from the inflationary one.

\section{Conclusions}

We have argued that for brane-antibrane inflation in strongly-warped
extra-dimensional vacua --- such as have been considered in detail
for Type IIB string models --- there is a natural mechanism which
channels a fraction of the released energy into reheating the
Standard Model degrees of freedom. This is because a nonnegligible
fraction of the false vacuum energy of the brane-antibrane system
naturally ends up being deposited into massless modes on branes which
are localized inside the most strongly-warped throats, rather than
being dumped completely into massless bulk-state modes.

This process relies on what is known about brane-antibrane
annihilation in flat space, where it is believed that the
annihilation energy dominantly produces very massive closed-string
states, which then quickly themselves decay to produce massive KK
modes for massless string states. What is important for our
purposes is that the wave functions for all of the massive KK
modes of this type are typically exponentially enhanced at the
bottom of warped throats, while those for the massless KK bulk
modes are not. This enhancement arises because the energies of
these states are minimized if their probabilities are greatest in
the most highly warped regions. This peaking is crucial because it
acts to suppress the couplings of the massive KK modes to the
massless bulk states, while enhancing their couplings to brane
modes in the most warped throats.

Although the couplings of the KK modes to SM degrees of freedom
are enhanced, the KK modes must first tunnel from the inflation
throat to the SM throat.  This results in most of the energy density
of the brane-antibrane system ending up as massless gravitons,
and only a small fraction $P$ going into visible matter.  
Nevertheless, for reasonable values of the string coupling and the
warp factor of the inflationary throat, $P$ can be as large as
$10^{-3} - 10^{-4}$.  Since the initial reheat temperature is high, there are
many decades of evolution in temperature during which the decoupled
gravitons can be diluted by events which increase the entropy of the
thermalized visible sector particles relative to the gravitons.
In this way it is quite plausible that big bang nucleosynthesis bounds
on the energy density of gravitons can be satisfied.

From this point of view, it is possible to efficiently reheat the SM
brane after brane-antibrane inflation, so long as there are no other
hidden branes lying in even deeper throats than the SM, which would
have a larger branching ratio for visible sector decay than the SM.
This observation is all the more interesting given the attention
which multiple-throat inflationary models are now receiving, both due
to the better understanding which they permit for the relation
between the inflationary scale and those of low-energy particle
physics, and to the prospects they raise for producing long-lived
cosmic string networks with potentially observable consequences.

\section*{Acknowledgements}
It is a pleasure to thank Shamit Kachru, Renata Kallosh, Andrei
Linde, Juan Maldacena, Anupam Mazumdar, Liam McAllister, Rob
Myers, Joe Polchinski, Fernando Quevedo, Raul Rabad\'an, 
Eva Silverstein, Horace
Stoica and Henry Tye for fruitful discussions. This research is
supported in part by funds from NSERC of Canada, FQRNT of Qu\'ebec
and McGill University.

\section{Appendix: The 4D View}

In this appendix we compute the low-energy couplings amongst the
bulk zero modes and brane modes in the effective 4D supergravity
obtained after modulus stabilization {\it \`a la} KKLT \cite{kklt}.
Besides checking the scaling of the kinetic terms obtained by
dimensionally reducing the Einstein-Hilbert action, this also
allows the study of the couplings in the scalar potential which
arise from modulus stabilization and so are more difficult to
analyze from a semiclassical, higher-dimensional point of view.

To this end imagine integrating out all of the extra-dimensional
physics to obtain the low-energy effective 4D supergravity for a
Type IIB GKP vacuum having only the mandatory volume modulus (and
its supersymmetric friends) plus various low-energy brane modes
(such as those describing the motion of various D3 branes). The
terms in this supergravity involving up to two derivatives are
completely described once the K\"ahler function, $K$,
superpotential, $W$, and gauge kinetic function, $f_{ab}$, are
specified.

Denoting the bulk-modulus supermultiplet by $T$ and the brane
multiplets by $\phi^I$, we use the K\"ahler potential
\cite{truncation,kklt,kklmmt,louis}
\begin{equation}
    K = -3 \log \left[ r \right] \,,
\end{equation}
where $r = T + T^* + k(\phi,\phi^*)$. For instance, if $\phi^I$
denotes the position of single brane, then $k$ is the K\"ahler
potential for the underlying 6D manifold. This implies the scalar
kinetic terms are governed by the following K\"ahler metric in
field space
\begin{equation}
    K_{TT^*} = \frac{3}{r^2} \,, \qquad
    K_{IT^*} = \frac{3 \, k_I}{r^2} \, \qquad \hbox{and} \qquad
    K_{IJ^*} = \frac{3}{r^2} \, \left[ k_I \, k_{J^*} - r\, k_{IJ^*}
    \right] \,,
\end{equation}
with inverse
\begin{equation}
    K^{T^*T} = \frac{r}{3} \, \left[ r - k^{L^*N} k_{L^*} k_N \right]
    \,, \quad
    K^{J^*T} = \frac{r\, k^{J^*L} k_L }{3} \quad \hbox{and} \quad
    K^{J^*I} = - \frac{r\, k^{J^*I}}{3}  \,.
\end{equation}

In the absence of modulus stabilization the superpotential of the
effective theory is a constant \cite{gvw}, $W = w_0$, and the
supergravity takes the usual no-scale form \cite{noscale}, with
vanishing scalar potential. If, however, there are low-energy
gauge multiplets associated with any of the D7 branes of the model
then their gauge kinetic function is $f_{ab} = T \, \delta_{ab}$.
For nonabelian multiplets of this type gaugino condensation
\cite{gc,ourgc} can generate a nontrivial superpotential, of the
form
\begin{equation}
    \qquad W = w_0 + A \, \exp \left[ - a \, T \right] \,,
\end{equation}
where $A$ and $a$ are calculable constants.

With these choices the K\"ahler derivatives of the superpotential
become
\begin{equation}
    D_T W = W_T - \frac{3 W}{r} \,, \qquad \hbox{and} \qquad
    D_I W = - \frac{3 k_I \, W}{r} \,,
\end{equation}
and so the supersymmetric scalar potential \cite{cremmeretal}
becomes
\begin{equation}
    V = \frac{1}{3r^2} \, \left[ \left(r - k^{I^*J} k_{I*} k_J
    \right) |W_T|^2 - 3 (W^* W_T + W W_T^*) \right] \,.
\end{equation}
Notice that use of these expression implicitly requires that we
work in the 4D Einstein frame, and so are using 4D Planck units
for which $M_p = {\cal O}(1)$.

If we specialize to the case of several branes, for which
$\{\phi^I \} = \{\phi^i_n \}$, with $i$ labelling the fields on a
given brane and $n = 1,\dots,N$ labelling which brane is involved,
then we typically have
\begin{equation}
    k(\phi^I,\phi^{I*}) = \sum_n k^{(n)}(\phi^i_n, \phi^{i*}_n)
    \,.
\end{equation}
In this case the K\"ahler metric built from $k$ is block diagonal,
with $k_{i_n j_m} = k^{(n)}_{ij} \, \delta_{mn}$, and so $k^{I^*J}
k_{I^*} k_J = \sum_n k_{(n)}^{i^* j} k^{(n)}_{i^*} k^{(n)}_{j}$
and so on.

We may now see how strongly the bulk KK zero modes, $g_{\mu\nu}$
and $T$, couple to one another and to the brane modes. Setting $k
= 0$ in the above shows that the couplings of $T$ and $g_{\mu\nu}$
to one another are order unity, and since our use of the standard
4D supergravity formalism requires us to be in the Einstein frame,
this implies these are all of 4D Planck strength (in agreement
with our higher-dimensional estimates).

Couplings to the branes are obtained by keeping $k$ nonzero, and
in the event that the branes are located in highly warped regions,
we must take $k^{(n)} = {\cal O}(a_n^2)$ with $a_n \ll 1$ denoting
the warp factor at the position of brane $n$.\footnote{For
instance, this power of $a_n$ reproduces the $a_n$-dependence of
the factor $\sqrt{g} g^{\mu\nu}$ obtained by dimensionally
reducing the higher-dimensional kinetic terms.} In this case the
combination $k_{(n)}^{i^* j} k^{(n)}_{i^*} k^{(n)}_{j}$ is also
${\cal O}(a_n^2)$.

Suppose we now expand the functions $k^{(n)}$ in powers of $\phi$
and keep only the leading powers:
\begin{equation}
    k^{(n)} \approx a_n^2 \sum_i \phi^{i*}_n \, \phi^i_n \,.
\end{equation}
Then, since the $\phi_n$ kinetic terms are ${\cal O}(a_n^2)$, we
see that the canonically-normalized fields are $\chi^i_n = a_n
\phi^i_n$. Once this is done the leading couplings to $T$ and
$g_{\mu\nu}$ are those which involve those parts of $k^{(n)}$ that
are quadratic in $\chi^i_n$, and since these are also order unity,
these couplings are also of Planck strength (again in agreement
with our earlier estimates).

Alternatively, consider now those couplings which only involve the
brane modes. Working to leading order in $a_n^2$, we see that a
term in $k^{(n)}$ of the form $(\chi^i_n)^k$ has a strength which
is of order $a_n^{2-k}$. For instance the case $k = 3$ generates
cubic couplings from the kinetic lagrangian of order $a_n^{-1}
\chi \partial \chi \partial \chi$, whose coefficient is of order
$(a_n M_p)^{-1} = M_{sm}^{-1}$. These are larger than Planck
suppressed ones, as expected.


\begin{thebibliography}{4}

\bibitem{SugraInflation}
D.V. Nanopoulos, K.A. Olive, M. Srednicki and K. Tamvakis, Phys.\
Lett.\ {\bf B123} (1983) 41;
%
A.B. Goncharov and A.D. Linde, Phys.\ Lett.\ {\bf B139} (1984) 27;
%
B. Gato, J. Leon, M. Quiros and M. Ramon-Medrano, Z.\ Phys.\ {\bf
C22} (1984) 345;
%
G. Gelmini, C. Kounnas and D.V. Nanopoulos, Nucl.\ Phys.\ {\bf
B250} (1985) 177;
%
L.G. Jensen and K.A. Olive, Nucl.\ Phys.\ {\bf B263} (1986) 731;
%
P.~Binetruy and M.~K.~Gaillard,
Phys.\ Rev.\ {\bf D34} (1986) 3069;
%
G.L. Cardoso and B.A. Ovrut, Phys.\ Lett.\ {\bf B298} (1993)
292-298 [hep-th/9210114];
%
R.~Brustein and P.~J.~Steinhardt,
Phys.\ Lett.\ {\bf B302}, 196 (1993) [hep-th/9212049];
%
J.A. Adams, G.G. Ross and S. Sarkar, Phys.\ Lett.\ {\bf B391}
(1997) 271-280 [hep-ph/9608336];
%
E. Halyo, Phys.\ Lett.\ {\bf B387} (1996) 43-47 [hep-ph/9606423];
%
P.~Binetruy and G.~R.~Dvali,
Phys.\ Lett.\ {\bf B388} (1996) 241 [hep-ph/9606342];
%
A.D. Linde and A. Riotto, Phys.\ Rev.\ {\bf D56} (1997) 1841-1844
[hep-ph/9703209].

\bibitem{DvaliTye}
G.~R.~Dvali and S.~H.~H.~Tye,
Phys.\ Lett.\ {\bf B450} (1999) 72 [hep-ph/9812483].

\bibitem{BI1}
C.~P.~Burgess, M.~Majumdar, D.~Nolte, F.~Quevedo, G.~Rajesh and
R.~J.~Zhang,
JHEP {\bf 0107} (2001) 047 [hep-th/0105204].

\bibitem{BBbarInflation}
S.~H.~Alexander,
Phys.\ Rev.\ {\bf D65} (2002) 023507 [hep-th/0105032];
%
G.~R.~Dvali, Q.~Shafi and S.~Solganik, ``D-brane inflation,''
[hep-th/0105203].

\bibitem{Angles}
J.~Garcia-Bellido, R.~Rabadan and F.~Zamora, ``Inflationary
scenarios from branes at angles,'' JHEP {\bf 0201} (2002) 036
[hep-th/0112147];
%
N.~Jones, H.~Stoica and S.~H.~Tye, ``Brane interaction as the
origin of inflation,'' JHEP {\bf 0207} (2002) 051
[hep-th/0203163];
%
M.~Gomez-Reino and I.~Zavala, ``Recombination of intersecting
D-branes and cosmological inflation,'' JHEP {\bf 0209} (2002) 020
[hep-th/0207278].

\bibitem{Others}
A. Mazumdar, S. Panda and A. P\'erez-Lorenzana, ``Assisted
Inflation via Tachyon Condensation,'' Nucl.\ Phys.\ {\bf B614}
(2001) 101-116 [hep-ph/0107058];
%
C.~P.~Burgess, P.~Martineau, F.~Quevedo, G.~Rajesh and
R.~J.~Zhang, ``Brane antibrane inflation in orbifold and
orientifold models,'' JHEP {\bf 0203} (2002) 052 [hep-th/0111025];
%
C.~Herdeiro, S.~Hirano and R.~Kallosh, ``String theory and hybrid
inflation / acceleration,'' JHEP {\bf 0112} (2001) 027
[hep-th/0110271];
%
K.~Dasgupta, C.~Herdeiro, S.~Hirano and R.~Kallosh, ``D3/D7
inflationary model and M-theory,'' Phys.\ Rev.\ D {\bf 65} (2002)
126002 [hep-th/0203019];
%
L.~Pilo, A.~Riotto and A.~Zaffaroni, ``Old inflation in string
theory,'' JHEP {\bf 0407}, 052 (2004)
[hep-th/0401004];
%
J.~P.~Hsu, R.~Kallosh and S.~Prokushkin, ``On brane inflation with
volume stabilization,'' JCAP {\bf 0312} (2003) 009
[hep-th/0311077];
%
F.~Koyama, Y.~Tachikawa and T.~Watari, ``Supergravity analysis of
hybrid inflation model from D3-D7 system'', [hep-th/0311191];
%

\bibitem{cosmicstrings}
S.~Sarangi and S.~H.~H.~Tye, ``Cosmic string production towards
the end of brane inflation,'' Phys.\ Lett.\ B {\bf 536} (2002) 185
[hep-th/0204074];
%
G.~Dvali, R.~Kallosh and A.~Van Proeyen, ``D-term strings,'' JHEP
{\bf 0401} (2004) 035 [hep-th/0312005];
%
G.~Dvali and A.~Vilenkin, ``Formation and evolution of cosmic
D-strings,'' JCAP {\bf 0403} (2004) 010 [hep-th/0312007].
%
J. Urrestilla, A. Achucarro and A.C. Davis, Phys.\ Rev.\ Lett.\
{\bf 92} (2004) 251302 [hep-th/0402032].

\bibitem{GKP}
S.~B.~Giddings, S.~Kachru and J.~Polchinski, ``Hierarchies from
fluxes in string compactifications,'' Phys. Rev. {\bf D66}, 106006
(2002).

\bibitem{Sethi}
S.~Sethi, C.~Vafa and E.~Witten, ``Constraints on low-dimensional
string compactifications,'' Nucl.\ Phys.\ B {\bf 480} (1996) 213
[hep-th/9606122];
%
K.~Dasgupta, G.~Rajesh and S.~Sethi, ``M theory, orientifolds and
G-flux,'' JHEP {\bf 9908} (1999) 023 [hep-th/9908088].

\bibitem{kklt}
  S.~Kachru, R.~Kallosh, A.~Linde and S.~P.~Trivedi,
``De Sitter vacua in string theory,''
  Phys.\ Rev.\ D {\bf 68}, 046005 (2003)
  [arXiv:hep-th/0301240].

\bibitem{kklmmt}
S.~Kachru, R.~Kallosh, A.~Linde, J.~Maldacena, L.~McAllister and
S.~P.~Trivedi, ``Towards inflation in string theory,'' JCAP {\bf
0310} (2003) 013 [hep-th/0308055].

\bibitem{racetrackinflation}
J.J.~ Blanco-Pillado, C.P.~Burgess, J.M.~Cline, C.~Escoda,
M.~G\'omez-Reino, R.~Kallosh, A.~Linde, and F.~Quevedo,
``Racetrack Inflation,'' 
  JHEP {\bf 0411}, 063 (2004)
  [arXiv:hep-th/0406230].


\bibitem{OtherStringInflation}
A.R. Frey, M. Lippert and B. Williams, ``The fall of stringy de
Sitter", Phys.\ Rev.\ {\bf D68} (2003) 046008 [hep-th/0305018];
%
 S.~Shandera, B.~Shlaer, H.~Stoica and S.~H.~H.~Tye,
  ``Inter-brane interactions in compact spaces and brane inflation,''
  JCAP {\bf 0402}, 013 (2004)
  [arXiv:hep-th/0311207].
%
C.~Escoda, M.~Gomez-Reino and F.~Quevedo, ``Saltatory de Sitter
string vacua,''JHEP {\bf 0311} (2003) 065, [hep-th/0307160];
%
C.~P.~Burgess, R.~Kallosh and F.~Quevedo, ``de Sitter string vacua
from supersymmetric D-terms,'' JHEP {\bf 0310} (2003) 056
[hep-th/0309187]
%
A.~Buchel and R.~Roiban, ``Inflation in warped geometries,''
Phys.\ Lett.\ B {\bf 590}, 284 (2004)
hep-th/0311154;
%
H.~Firouzjahi and S.~H.~H.~Tye, ``Closer towards inflation in
string theory,''
 Phys.\ Lett.\ B {\bf 584}, 147 (2004)
 [hep-th/0312020];
%
A. Saltman and E. Silverstein, ``The Scaling of the No Scale
Potential and de Sitter Model Building,'' 
  JHEP {\bf 0411}, 066 (2004)
[hep-th/0402135];
%
E.~Halyo, ``D-brane inflation on conifolds,'' hep-th/0402155;
%
R.~Kallosh and S.~Prokushkin, ``SuperCosmology,''
[hep-th/0403060];
%
M. Becker, G. Curio and A. Krause, Nucl.\ Phys.\ {\bf B693} (2004)
223-260 [hep-th/0403027];
%
 M.~Alishahiha, E.~Silverstein and D.~Tong,
``DBI in the sky,''
  Phys.\ Rev.\ D {\bf 70}, 123505 (2004)
  [arXiv:hep-th/0404084].
%
J.~P.~Hsu and R.~Kallosh, ``Volume stabilization and the origin of
the inflaton shift symmetry in string theory,'' JHEP {\bf 0404}
(2004) 042 [hep-th/0402047];
%
 O.~DeWolfe, S.~Kachru and H.~Verlinde,
``The giant inflaton,'' JHEP {\bf 0405} (2004) 017
[hep-th/0403123].

\bibitem{BIReviews}
F.~Quevedo, Class.\ Quant.\ Grav.\  {\bf 19} (2002) 5721,
[hep-th/0210292];
%
A.~Linde, ``Prospects of inflation,'' [hep-th/0402051];
%
C.P. Burgess, 
``Inflatable string theory?,''
  Pramana {\bf 63}, 1269 (2004)
  [arXiv:hep-th/0408037].


\bibitem{twothroat}
N.~Iizuka and S.~P.~Trivedi, ``An inflationary model in string
theory,'' 
 Phys.\ Rev.\ D {\bf 70}, 043519 (2004)
hep-th/0403203.
%
X. Chen, ``Multi-throat Brane Inflation,'' [hep-th/040084].

\bibitem{stringycosmicstrings}
E.~J.~Copeland, R.~C.~Myers and J.~Polchinski, ``Cosmic F- and
D-strings,'' JHEP {\bf 0406} (2004) 013 [hep-th/0312067];
%
L.~Leblond and S.~H.~H.~Tye, ``Stability of D1-strings inside a
D3-brane,'' JHEP {\bf 0403} (2004) 055 [hep-th/0402072];
%
K.~Dasgupta, J.~P.~Hsu, R.~Kallosh, A.~Linde and M.~Zagermann,
``D3/D7 brane inflation and semilocal strings,'' (hep-th/0405247);
%
L. Leblond and S.H. Tye, ``Stability of D1 Strings Inside a D3
Brane,'' JHEP 0403 (2004) 055 [hep-th/0402072].

\bibitem{cmqu}
J.~F.~G.~Cascales, M.~P.~Garcia del Moral, F.~Quevedo and
A.~M.~Uranga, ``Realistic D-brane models on warped throats:
Fluxes, hierarchies and moduli stabilization,'' hep-th/0312051.

\bibitem{bcqs}
C.~P.~Burgess, J.~M.~Cline, H.~Stoica and F.~Quevedo, ``Inflation
in realistic D-brane models,'' [hep-th/0403119].

\bibitem{RA}
R. Kallosh and A. Linde,
``Landscape, the scale of SUSY breaking, and inflation,''
  JHEP {\bf 0412}, 004 (2004)
  [arXiv:hep-th/0411011].

\bibitem{braneheat}
Y. Himemoto and T. Tanaka, Phys.\ Rev.\ {\bf D67} (2003) 084014
[gr-qc/0212114];
%
R. Allahverdi, A. Mazumdar and A. Perez-Lorenzana, Phys.\ Lett.\
{\bf B516} (2001) 431-438 [hep-ph/0105125];
%
J.H. Brodie and D.A. Easson, JCAP 0312 (2003) 004
[hep-th/0301138];
%
Y. Takamizu and K. Maeda, ``Collision of Domain Walls and
Reheating of the Brane Universe,'' [hep-th/0406235];
%
E. Papantonopoulos and V. Zamarias, JHEP 0410 (2004) 051
[hep-th/0408227];
%
C.J. Chang and I.G. Moss, ``Brane Inflation With Dark Reheating,''
[hep-ph/0411021].

\bibitem{RS}
L.~Randall and R.~Sundrum,
``A large mass hierarchy from a small extra dimension,''
Phys.\ Rev. Lett. {\bf 83} ({1999}) {3370} [\hepph{9905221}];
%
L.~Randall and R.~Sundrum,
``An alternative to compactification,''
Phys. Rev. Lett. {\bf 83} ({1999}) {4690} [\hepth{9906064}].

\bibitem{IntScale}
K.~Benakli, {\it Phenomenology of low quantum gravity scale
models,} Phys.\ Rev.\ D {\bf 60}, 104002 (1999) [hep-ph/9809582];
C.~P.~Burgess, L.~E.~Ibanez and F.~Quevedo, {\it Strings at the
intermediate scale or is the Fermi scale dual to the  Planck
scale?} Phys.\ Lett.\ B {\bf 447}, 257 (1999) [hep-ph/9810535].

\bibitem{Rob}
A. Frey, R. Myers and A. Mazumdar, in preparation.

\bibitem{throatmetric}
O. De Wolfe and S.B. Giddings, ``Scales and Hierarchies in Warped
Compactifications and Brane Worlds,'' Phys.\ Rev.\ {\bf D67}
(2003) 066008 [hep-th/0208123].

\bibitem{dkkls}
 S.~Dimopoulos, S.~Kachru, N.~Kaloper, A.~E.~Lawrence and E.~Silverstein,
  ``Generating small numbers by tunneling in multi-throat  compactifications,''
  Int.\ J.\ Mod.\ Phys.\ A {\bf 19}, 2657 (2004)
  [arXiv:hep-th/0106128];
S.~Dimopoulos, S.~Kachru, N.~Kaloper, A.~E.~Lawrence and E.~Silverstein,
  ``Small numbers from tunneling between brane throats,''
  Phys.\ Rev.\ D {\bf 64}, 121702 (2001)
  [arXiv:hep-th/0104239].

\bibitem{PlanckBraneModes}
W.~D.~Goldberger and M.~B.~Wise, ``Bulk fields in the
Randall-Sundrum compactification scenario,'' Phys.\ Rev.\ D {\bf
60}, 107505 (1999) [hep-ph/9907218];
%
H.~Davoudiasl, J.~L.~Hewett and T.~G.~Rizzo, ``Phenomenology of
the Randall-Sundrum gauge hierarchy model,'' Phys.\ Rev.\ Lett.\
{\bf 84}, 2080 (2000) [hep-ph/9909255].

\bibitem{BraneRad}
L. McAllister and I. Mitra, ``Relativistic D-Brane Scattering is
Extremely Inelastic,'' [hep-th/0408085].

\bibitem{Sen}
A. Sen, ``Tachyon Matter,'' JHEP 0207 (2002) 065 [hep-th/0203265].

\bibitem{BraneDecay}
A. Sen,
``Rolling Tachyon,''
JHEP {\bf 0204} (2002) 048 [hep-th/0203211];
%
N. Lambert, H. Liu and J. Maldacena,
[hep-th/0303139].

\bibitem{Sen-review}
A.~Sen, ``Tachyon dynamics in open string theory,''
hep-th/0410103.

\bibitem{HiJCascade}
R. Iengo and J.G. Russo,
JHEP {\bf 0211} (2002) 045 [hep-th/0210245];
%
R. Iengo and J.G. Russo,
JHEP {\bf 0303} (2003) 030 [hep-th/0301109];
%
D. Chialva, R. Iengo and J.G. Russo,
JHEP {\bf 0312} (2003) 014 [hep-th/0310283];
%
D. Chialva, R. Iengo and J.G. Russo,
[hep-th/0410152].

\bibitem{RobnJoe}
J. Lykken, R. Myers and J. Wang, ``Gravity in a Box,'' JHEP {\bf
0009} (2000)009 [hep-th/0006191].


\bibitem{Reheat}
L.~F.~Abbott, E.~Farhi and M.~B.~Wise,
``Particle Production In The New Inflationary Cosmology,''
Phys.\ Lett.\ B {\bf 117}, 29 (1982);
A.~D.~Linde,
``Particle Physics And Inflationary Cosmology,''
 Chur, Switzerland: Harwood (1990) 362 p. (Contemporary concepts in physics, 5).

\bibitem{truncation}
E.~Witten, ``Dimensional Reduction Of Superstring Models,'' Phys.\
Lett.\ B {\bf 155} (1985) 151;
%
C.~P.~Burgess, A.~Font and F.~Quevedo, ``Low-Energy Effective
Action For The Superstring,'' Nucl.\ Phys.\ B {\bf 272} (1986)
661.

\bibitem{louis}
M. Gra\~a, T.W. Grimm, H. Jockers and J. Louis, Nucl.\ Phys.\ {\bf
B690} (2004) 21-61 [hep-th/0312232];
%
T.W. Grimm and J. Louis, Nucl.\ Phys.\ {\bf B699} (2004) 387-426
[hep-th/0403067].

\bibitem{gvw}
S. Gukov, C. Vafa and E. Witten, ``CFTs from Calabi-Yau
Fourfolds,'' Nucl. Phys. {\bf B584}, 69 (2000).

\bibitem{noscale}
E. Cremmer, S. Ferrara, C. Kounnas and D.V. Nanonpoulos,
``Naturally vanishing cosmological constant in $N=1$
supergravity,'' Phys. Lett. {\bf B133}, 61 (1983);
%
J. Ellis, A.B. Lahanas, D.V. Nanopoulos and K. Tamvakis,
``No-scale Supersymmetric Standard Model,'' Phys. Lett. {\bf
B134}, 429 (1984).

\bibitem{gc}
J.~P.~Derendinger, L.~E.~Ibanez and H.~P.~Nilles, ``On The
Low-Energy D = 4, N=1 Supergravity Theory Extracted From The D =
10, N=1 Superstring,'' Phys.\ Lett.\ B {\bf 155} (1985) 65;
%
M.~Dine, R.~Rohm, N.~Seiberg and E.~Witten, ``Gluino Condensation
In Superstring Models,'' Phys.\ Lett.\ B {\bf 156} (1985) 55.

\bibitem{ourgc}
C.P. Burgess, J.-P. Derendinger, F. Quevedo and M. Quir\'os,
``Gaugino Condensates and Chiral-Linear Duality: An
Effective-Lagrangian Analysis'', Phys.\ Lett.\ B {\bf 348} (1995)
428--442;
``On Gaugino Condensation with Field-Dependent Gauge Couplings'',
Ann.\ Phys.\ {\bf 250} (1996) 193-233.

\bibitem{cremmeretal}
E.~Cremmer, B.~Julia, J.~Scherk, S.~Ferrara, L.~Girardello and
P.~van Nieuwenhuizen, ``Spontaneous Symmetry Breaking And Higgs
Effect In Supergravity Without Cosmological Constant,'' Nucl.\
Phys.\ B {\bf 147}, 105 (1979).

\bibitem{KKM}
M.~Kawasaki, K.~Kohri and T.~Moroi,
``Hadronic decay of the gravitino in the early universe and its implications
to inflation,''
arXiv:hep-ph/0410287.


\bibitem{splitsusy}
N.~Arkani-Hamed and S.~Dimopoulos,
``Supersymmetric unification without low energy supersymmetry and signatures
for fine-tuning at the LHC,''
arXiv:hep-th/0405159;

G.~F.~Giudice and A.~Romanino,
``Split supersymmetry,''
Nucl.\ Phys.\ B {\bf 699}, 65 (2004)
[arXiv:hep-ph/0406088].



\end{thebibliography}
\end{document}